\documentclass[twocolumn,prl,preprintnumbers,
nofootinbib,amsmath]{revtex4}

\allowdisplaybreaks

\renewcommand{\d}{\mathrm{d}}
\newcommand{\I}{\mathrm{i}}
\newcommand{\al}{\alpha}
\newcommand{\de}{\delta}
\newcommand{\la}{\lambda}
\newcommand{\e}{\mathrm{e}}
\newcommand{\cA}{\mathcal{A}}
\newcommand{\cF}{\mathcal{F}}
\newcommand{\cG}{\mathcal{G}}
\newcommand{\cM}{\mathcal{M}}
\newcommand{\cN}{\mathcal{N}}
\newcommand{\bD}{\bar{D}}
\newcommand{\one}{\openone}  

\newcommand{\half}{\tfrac{1}{2}}
\newcommand{\f}[2]{f_{#1}{}^{#2}}
\newcommand{\frc}[2]{\frac{\raisebox{-1pt}{$#1$}}{#2}}

\begin{document} 

\preprint{FSU-TPI-11/04}

\title{Masses and Dualities in Extended Freedman-Townsend Models}

\author{Ulrich Theis}

\affiliation{%
Institute for Theoretical Physics \\
Friedrich-Schiller-University Jena \\
Max-Wien-Platz 1, D-07743 Jena, Germany}

\email{Ulrich.Theis@uni-jena.de}

\begin{abstract}

We consider some generalizations of Freedman-Townsend models of
self-interacting antisymmetric tensors, involving couplings to further
form fields introduced by Henneaux and Knaepen. We show how these fields
can provide masses to the tensors by means of the St\"uckelberg
mechanism and implement the latter in four-dimensional $\cN=1$
superspace. The duality properties of the form fields are studied, and
the paradoxical situation of a duality between a free and an interacting
theory is encountered.

\end{abstract}



\maketitle

\section{Introduction} 

In the effort to relate Calabi-Yau compactifications of ten-dimensional
superstring theories to the real world, the method of turning on
electric and magnetic fluxes of various $p$-form fields has gained a lot
of attention due to the observation that the resulting scalar potentials
in the low-energy effective supergravity theories can lift the vacuum
degeneracy.

In four-dimensional models derived from such flux compactifications of
Type II strings, massive antisymmetric tensor fields (2-forms) appear
\cite{LM}, whose properties until recently have not been investigated in
much detail. This situation is quickly improving, however; in
particular, the corresponding gauged $\cN=2$ supergravity theories
containing (multi-) tensor multiplets have been worked out in
\cite{DDSV,DSV}, drawing on results obtained in \cite{TV}.

The simultaneous introduction of electric and magnetic charges in this
context still poses some problems. Recently, D'Auria et al.\ have
suggested in \cite{DFTV} that a reformulation in terms of
Freedman-Townsend (FT) models may provide a solution. These models
describe $(n-2)$-forms in $n$ dimensions with non-polynomial
interactions and are dual to non-linear sigma models of scalar fields
\cite{OP,FT}. In flux compactifications of Type II strings, FT models
can arise by dualization of the Ramond-Ramond scalars into tensors; the
non-abelian nature of the (Heisenberg) algebra of isometries of the
target space metric translates into FT-like interactions. The question
then arose in \cite{DFTV} how to couple the self-interacting tensors to
vector fields in a way which corresponds to turning on RR fluxes.
Magnetic charges in particular should introduce mass terms for the
tensors in a gauge invariant manner. This was achieved for the
Neveu-Schwarz 2-form by means of the St\"uckelberg mechanism, but not
for the remaining RR tensors. In this letter we close this gap by
deriving the most general massive FT model.

Toward this end, we employ an extension of FT models constructed by
Henneaux and Knaepen in \cite{HK1}, which involves couplings of the
$(n-2)$-forms to $p$-forms of lower rank. These additional fields will
provide the degrees of freedom that turn the tensors into massive
fields. Gauge invariance is maintained through the St\"uckelberg
mechanism as in \cite{DFTV}. This is different from the methods already
described by Freedman and Townsend \cite{FT}, where either gauge
invariance was broken by the mass terms, or where through a coupling to
Yang-Mills fields the latter were made massive.

Our purely bosonic models can be generalized to $\cN=1$ supersymmetric
models using the superfield approach of \cite{BT}, as we demonstrate in
the last section of this letter. A massive $\cN=1$ tensor multiplet was
first derived in \cite{S}, and has recently been formulated in \cite{GL}
(see also \cite{DF,LS,K}) using the St\"uckelberg mechanism. Here we
generalize the latter construction to an arbitrary number of tensor
multiplets with self-interactions of the FT-type.

Before we turn to the construction of massive FT models, we first review
the Henneaux-Knaepen (HK) extensions and make some interesting
observations about their duality properties, which to our knowledge have
not been presented in the literature so far.

\section{Dualities in extended FT models} 

Consider sets of $(n-2)$-forms $B_I$, $p$-forms $A^a$ with $p<n-2$, and
auxiliary 1-forms $V^I$ in $n>2$ dimensions. The HK action in
first-order form reads \cite{HK1} (where we write $A^a$ as a column
vector and wedge products are understood)
 \begin{align} \label{S_1}
  S_1 = \frc{1}{2} \int\! \big( & \Delta_{IJ} *\! V^I\, V^J + 2 B_I G^I
	\notag \\[-3pt]
  & + *\cF^t M \cF + \cF^t N \cF \big)\ ,
 \end{align}
with field strengths
 \begin{align}
  G^I & = \d V^I + \half\, \f{JK}{I} V^J V^K \\[2pt]
  \cF & = (\d + V^I T_I) A\ . \label{cF}
 \end{align}
The $\f{IJ}{K}$ are structure constants of some arbitrary Lie algebra
$\cG$, while the real matrices ${{T_I}^a}_b$ define a representation of
$\cG$,
 \begin{equation}
  \f{[IJ}{L} \f{K]L}{M} = 0\ ,\quad [\, T_I\, ,\, T_J\, ] = \f{IJ}{K}
  T_K\ .
 \end{equation}
$\Delta$ and $M$ are arbitrary positive-definite symmetric matrices. The
topological term proportional to the matrix $N=(-)^{p+1}N^t$ can be
present only if $2(p+1)=n$. It was not part of the original formulation
of HK models, but should be included as it arises naturally in
supersymmetric extensions \cite{BT}. $\Delta$, $M$, and $N$ are neither
required to be $\cG$-invariant, nor do they have to be constant and can
depend on further fields.

$S_1$ is invariant under abelian gauge transformations $\de=\de_\la+
\de_C$ with
 \begin{align}
  \de A & = (\d + V^I T_I) \la \quad\Rightarrow\quad \de \cF = G^I T_I
	\la \\[2pt]
  \de B_I & = \d C_I + \f{IJ}{K} V^J C_K - (*\cF^t M + \cF^t N)\, T_I
	\la \label{deB} \\[2pt]
  \de V^I & = 0\ ,
 \end{align}
where $\la^a$ and $C_I$ are arbitrary $(p-1)$- and $(n-3)$-forms,
respectively. Invariance under $\de_\la$ is obvious, while invariance
under $\de_C$ (modulo a boundary term) follows from the identity $\d
G^I+\f{JK}{I}V^J G^K=0$. Note that, despite of the similarity of $G^I$
to a Yang-Mills field strength, there is no gauge transformation
associated with $V^I$.

The second-order version is obtained by eliminating $V^I$ via their
algebraic equations of motion, which can be brought into the form
 \begin{align} \label{V=dB}
  & \big[ (\Delta_{IJ} + A^t M_{IJ}\, i_A) * - \f{IJ}{K} B_K - (-)^p
	A^t N_{IJ} A \big] V^J = \notag \\[2pt]
  & = (-)^n\, \d B_I - (-)^p\, (* F^t M + F^t N)\, T_I A\ ,
 \end{align}
where $F=\d A$. Here we have introduced the matrices
 \begin{equation}
  M_{IJ} = {T_I}^t M\, T_J\ ,\quad N_{IJ} = {T_I}^t N\, T_J\ ,
 \end{equation}
and the operator
 \begin{equation}
  i_{A_p} = \frc{1}{p!}\, A^{\mu_1\dots\mu_p} \frc{\partial^p}{\partial
  \d x^{\mu_p} \dots \partial \d x^{\mu_1}}\ ,
 \end{equation}
which arises in the above equation by virtue of the identity
$*(\omega_q A_p)=(-)^{p(n-1)}\,i_{A_p}\!*\!\omega_q$. In general,
\eqref{V=dB} can be solved for $V^I$ only perturbatively, the solution
being non-polynomial in both $B_I$ and $A^a$. Some examples of exact
solutions (involving one $B$ and one $A$) can be found in
\cite{BD,BT,T}. Note that when $N$-terms are present, they give rise
to couplings of $\d B_I$ to abelian Chern-Simons forms of $A^a$.

To derive the dual formulation one solves the equations of motion of
$B_I$, namely the flatness conditions $G^I=0$, in terms of scalars
$\phi^I$:
 \begin{equation}
  V^I T_I = \e^{-\phi\cdot T} \d \e^{\phi\cdot T} = \d\phi^I
  {E_I}^J(\phi)\, T_J\ .
 \end{equation}
The vielbein ${E_I}^J$ can be expressed as
 \begin{equation}
  {E_I}^J(\phi) = \int_0^1\! \d t\, (\e^{t \phi\,\cdot f})_I{}^J\
  ,\quad (\phi\cdot f)_I{}^J = \phi^K \f{IK}{J}\ .
 \end{equation}
When inserted into $\cF$, a local field redefinition allows to absorb
the flat connections $V^I(\phi)$,
 \begin{equation}
  \cF = \e^{-\phi\cdot T} \hat{F}\ ,\quad \hat{F} = \d \hat{A}\
  ,\quad \hat{A} = \e^{\phi\cdot T}\! A\ .
 \end{equation}
The dual action in terms of the scalars $\phi^I$ and the redefined
$p$-forms $\hat{A}$ then reads
 \begin{equation}
  S_2 = \frc{1}{2} \int\! \big( g_{IJ} *\! \d\phi^I\, \d\phi^J +
  *\hat{F}^t \hat{M} \hat{F} + \hat{F}^t \hat{N} \hat{F} \big)\ .
 \end{equation}
Here the target space metric of the NLSM is given by
 \begin{equation}
  g_{IJ}(\phi) = {E_I}^K(\phi)\, {E_J}^L(\phi)\, \Delta_{KL}\ ,
 \end{equation}
while the gauge coupling functions are
 \begin{equation} \label{hatMN}
  \hat{M}(\phi) = \e^{-\phi\cdot T^t}\! M \e^{-\phi\cdot T}\ ,\quad
  \hat{N}(\phi) = \e^{-\phi\cdot T^t}\! N \e^{-\phi\cdot T}\ .
 \end{equation}

It is now obvious (although it was not observed in \cite{HK1}) that
also the $p$-forms can be dualized, since the $\hat{A}$ enter the
action only via their derivatives. By adding to $S_2$ a term
$\d\hat{A}_{\tilde{p}}{}^t\,\hat{F}$ (where $\tilde{p}=n-2-p$) and
integrating out $\hat{F}$, we arrive at the dual action for the
$\tilde{p}$-forms $\hat{A}_{\tilde{p}}$. In the process, the matrices
$\hat{M}$ and $\hat{N}$ get inverted. From \eqref{hatMN} it is clear
that this effectively interchanges $T_I$ and $-{T_I}^t$. We can now
reverse the whole procedure; a redefinition $\hat{A}_{\tilde{p}}=
\e^{-\phi\cdot T^t}\!A_{\tilde{p}}$ reintroduces flat connections
$V^I(\phi)$, which can be promoted to unconstrained fields by enforcing
the flatness conditions $G^I=0$ through Lagrange multipliers $B_I$.
This results in a first-order action of the form \eqref{S_1} for $B_I$
and $A_{\tilde{p}\,a}$. We conclude that HK models with $p$-forms and
$\tilde{p}$-forms, respectively, are dual to each other upon the
substitution
 \begin{align}
  T_I & \leftrightarrow - {T_I}^t \notag \\[2pt]
  M & \leftrightarrow \big[ M - (-)^p N M^{-1} N \big]^{-1}
	\notag \\[2pt]
  N & \leftrightarrow (-)^p \big[ N - (-)^p M N^{-1} M \big]^{-1}\ .
 \end{align}
Note that the matrices $-{T_I}^t$ form a representation of the Lie
algebra $\cG$.

Let us repeat the steps we have taken: First dualize the $B_I$ into
$\phi^I$, then redefine $A_p\rightarrow\hat{A}_p$, dualize these
$\hat{A}_p$ into $\hat{A}_{\tilde{p}}$, and after another redefinition
$\hat{A}_{\tilde{p}}\rightarrow A_{\tilde{p}}$ dualize the $\phi^I$ back
into $B_I$. This raises the question whether there is a more direct
dualization from one HK model to the other.

A puzzling observation is that if $(MT_I)^t=-MT_I$ and $(NT_I)^t=(-)^p
NT_I$ for all $I$ (for example antisymmetric $T_I$ in the case $M=\one$
and $N=0$, which corresponds to the original HK models), then $\hat{M}
=M$, $\hat{N}=N$, and the $\hat{A}^a$ decouple from the $\phi^I$,
 \begin{equation} \label{S2free}
  S_2 = \frc{1}{2} \int\! \big( g_{IJ} *\! \d\phi^I\, \d\phi^J +
  *\hat{F}^t M \hat{F} + \hat{F}^t N \hat{F} \big)\ .
 \end{equation}
If one now dualizes the $\phi^I$ back into tensors $B_I$, no
interactions between the latter and $\hat{A}^a$ arise. Indeed,
\eqref{S2free} can be obtained from a first-order action without
couplings of $\hat{A}^a$ to $V^I$, 
 \begin{align}
  S_1 = \frc{1}{2} \int\! \big( & \Delta_{IJ} *\! V^I\, V^J + 2 B_I
	G^I \notag \\*[-3pt]
  & + *\hat{F}^t M \hat{F} + \hat{F}^t N \hat{F} \big)\ ,
 \end{align}
hence after elimination of $V^I$ the $B_I$ and $\hat{A}^a$ do not
interact.

This seems to provide a duality between an interacting theory (coupling
fields $B_I$ and $A^a$) and a free theory (with decoupled fields
$\phi^I$ and $\hat{A}^a$), disregarding self-interactions of the $B_I$
and $\phi^I$, respectively, for the sake of argument\footnote{In the
simplest setup where this phenomenon occurs, namely one $B$ and two
$A^a$ with ${T^a}_b=\varepsilon_{ab}$, $M=1$, $N=0$, there are no
interactions at all on the scalar side of the duality.}. We are not
aware of a resolution of this paradox. That the tensor side of the
duality really involves genuine interactions between $B_I$ and $A^a$
should be ensured by the fact that their couplings cannot be removed by
any local field redefinition. This was proved in \cite{HK2} by BRST
cohomological methods for any representation of $\cG$. The standard
duality procedure we have employed is a non-local operation, of course,
and under the specific circumstances spelled out above it can be
followed with a local field redefintion such that the combination
apparently amounts to a non-local field redefinition on the other side
of the duality which removes the interactions.

\section{Massive FT models} 

Let us now specialize to the case $p=n-3$. Then it is possible to
introduce masses for the tensors $B_I$ of pure FT models by means of
the St\"uckelberg mechanism, where the $p$-forms $A^a$ serve as
compensators. Toward this end we introduce constant mass parameters
$m^{aI}$ and another set of parameters $n_{Ia}(m)$, which are assumed
to satisfy the orthonormality condition
 \begin{equation} \label{nm}
  n_{Ia} m^{aJ} = \de_I^J\ .
 \end{equation}
Note that one needs at least as many $p$-forms as there are tensors,
$(n_A=\de_a^a)\geq(n_B=\de_I^I)$. Moreover, if $n_A>n_B$, the $n_{Ia}$
are in general not uniquely defined. Condition \eqref{nm} guarantees
that the matrices
 \begin{equation} \label{Tmn}
  {{T_I}^a}_b = m^{aJ}\! \f{JI}{K} n_{Kb}
 \end{equation}
form a representation of the Lie algebra $\cG$ with structure constants
$\f{IJ}{K}$.
The gauge transformations
 \begin{gather}
  \de A^a = \d \la^a + m^{aI} C_I\ ,\quad \de V^I = 0 \label{deA}
	\\[2pt]
  \de B_I = \d C_I + \f{IJ}{K} V^J (C_K + n_{Ka} \d \la^a) \label{deBm}
 \end{gather}
leave invariant the combination $\cF^a-m^{aI}B_I$ with $\cF^a$ as in
\eqref{cF} and $T_I$ as in \eqref{Tmn},
 \begin{equation}
  \de \cF^a = m^{aI} \de B_I\ .
 \end{equation}
Since the tensor transformations \eqref{deBm} are of the form
\eqref{deB} with $C_I=n_{Ia}\de A^a$ and $\la^a=0$, it immediately
follows that the first-order action
 \begin{align}
  S_1 = \frc{1}{2} \int\! \big[ & \Delta_{IJ} *\! V^I\, V^J + 2 B_I G^I
	\notag \\[-4pt]
  & + M_{ab} *\! (\cF^a - m^{aI} B_I)\, (\cF^b - m^{bJ} B_J) \notag
	\\[2pt]
  & + N_{ab}\, (\cF^a - m^{aI} B_I)\, (\cF^b - m^{bJ} B_J) \big]
 \end{align}
is gauge invariant. The topological terms in the last line are now only
possible if $n=4$ and $p=1$, which is just what we need in applications
to flux compactifications as in \cite{DFTV}. The auxiliary fields $V^I$
can again be eliminated; their equations of motion are obtained from
\eqref{V=dB} by replacing $F^a$ with $F^a-m^{aI}B_I$ and $T_I$ with
\eqref{Tmn}.

Since $S_1$ depends on the inverse mass parameters $n_{Ia}$ through the
mass ratios $T_I$, taking the massless limit is subtle: The projector
 \begin{equation} \label{P}
  {P^a}_b = \de^a_b - m^{aI} n_{Ib}
 \end{equation}
has $n_B$ eigenvectors $m^{aI}$ with zero eigenvalue, and we can
use the freedom in solving \eqref{nm} for $n_{Ia}$ to choose the latter
such that $P$ has $n_B$ vanishing rows. With this choice, we introduce
rescaled fields
 \begin{gather}
  \cA_I = n_{Ia} A^a\ ,\quad \cA^a = {P^a}_b A^b \notag \\[2pt]
  \Leftrightarrow \quad A^a = m^{aI}\! \cA_I + \cA^a\ .
 \end{gather}
$\cA^a$ has $n_A-n_B$ non-vanishing components. When expressed in terms
of these fields, the inverse mass parameters disappear from the action
due to ${T_I}^a{}_b A^b=m^{aJ}f_{JI}{}^K\cA_K$, and the massless limit
can be safely taken. The $\cA_I$ are precisely the compensators that
are eaten by $B_I$ in the ``unitary gauge'' where the $\cA_I$ are
transformed away using \eqref{deA}.

Note that if we define ``electric charges'' ${e_a}^I=N^{(0)}_{ab}
m^{bI}$, where the superscript $(0)$ refers to the constant part of
$N_{ab}$, then
 \begin{equation} \label{me}
  m^{aI} {e_a}^J - m^{aJ} {e_a}^I = 0
 \end{equation}
trivially by virtue of the symmetry of $N_{ab}^{(0)}$.

The mass matrix for the $B_I$ is complicated in general. To determine
it we can drop all interactions, which amounts to setting $\f{IJ}{K}=0$
and $\Delta_{IJ}^{\!(0)}\,V^J=*\d B_I$ in $S_1$. We then follow D'Auria
and Ferrara \cite{DF} in deriving the linearized equations of motion
for the fields
 \begin{equation}
  b_I = (B - \cM^{-1} \cN\! *\! B)_I\ ,
 \end{equation}
which are transversal on-shell, $\d\!*\!b_I=0$. Here we have defined
 \begin{equation}
  \cM^{IJ} = m^{aI} M^{(0)}_{ab} m^{bJ}\ ,\quad \cN^{IJ} = m^{aI}
  {e_a}^J
 \end{equation}
and assumed that $\cM$ is invertible. A short calculation yields the
equations
 \begin{equation}
  \big[\! *\!\d *\!\d - \big( \one + \cM^{-1} \cN \cM^{-1} \cN \big)
  \Delta^{\!(0)} \cM \big]{}_I{}^J\, b_J = 0\ ,
 \end{equation}
from which we can read off the mass matrix. As is familiar from flux
compactifications, it receives contributions from the electric charges.

The tensors $B_I$ could be eliminated instead of the $V^I$, which then
gives rise to a theory of massive vectors.

\section{Massive FT models in superspace} 

Just like the massless FT models \cite{LR,CLL,FINN} and their HK
extensions \cite{BT}, the massive four-dimensional FT models can be
formulated in $\cN=1$ superspace. Following \cite{BT}, we embed the
tensors $B_I$ in chiral spinor superfields $\Psi_{\al I}$ and the
vectors $A^a$, $V^I$ in real scalar superfields, for which we shall use
the same labels. For the latter, we construct super-field strengths
 \begin{align}
  Y_\al^a & = - \frc{\I}{4}\, \bD^2 \big[ \e^{-2\I V\cdot T} D_\al
	(\e^{\I V\cdot T}\! A) \big]^a \\[2pt]
  W_{\!\al}^I T_I & = - \frc{\I}{4}\, \bD^2 \big( \e^{-2\I V\cdot T}
	D_\al \e^{2\I V\cdot T} \big)\ ,
 \end{align}
with $T_I$ as in \eqref{Tmn}. These being chiral, the first-order action
 \begin{align}
  S_1 = \int\! \d^4x\, \d^2\theta\, \big[ & k_{ab}\, (Y^a - m^{aI}
	\Psi_I) (Y^b - m^{bJ} \Psi_J) \notag \\[-3pt]
  & + W^I \Psi_I + \d^2\bar\theta\, h(V) \big] + \text{c.c.}\ ,
	\label{ssS}
 \end{align}
where $k_{ab}=k_{ba}$ may depend on further chiral superfields, is
manifestly supersymmetric. The function $h(V)$ need not be quadratic
in $V^I$, as long as the latter can be eliminated.

The nontrivial part is to verify gauge invariance. Let us start with
the simpler tensor gauge transformations; embedding the parameters
$C_I$ into real scalar superfields of the same name, these read
 \begin{gather}
  \de_C A^a = m^{aI} C_I\ ,\quad \de_C V^I = 0 \\[2pt]
  \de_C \Psi_{\al I} = - \frc{\I}{4}\, \bD^2 \big[ \e^{-2\I V\cdot f}
	D_\al (\e^{\I V\cdot f} C) \big]_I\ .
 \end{gather}
Here we use $(V\!\cdot f)_I{}^J=V^K\f{IK}{J}$. To compute the induced
transformations of $Y^a$, we need the identity
 \begin{equation} \label{eiVT}
  (\e^{\I V\cdot T})^a{}_b = {P^a}_b + m^{aI} (\e^{\I V\cdot f})_I{}^J
  n_{Jb}\ ,
 \end{equation}
which follows from $(V\!\cdot T)^a{}_b=m^{aI}(V\!\cdot f)_I{}^J n_{Jb}$
and \eqref{nm} ($P$ was defined in \eqref{P}). It is now easy to show
that $\de_CY^a=m^{aI}\de_C\Psi_I$, which implies invariance of the first
line in \eqref{ssS}; just move $m^{aI}$ in $\de_C A^a$ past the $\e^{\I
V\cdot T}$'s in $Y^a$, which get converted into $\e^{\I V\cdot f}$'s. A
proof of invariance of the second line can be found in section 3 of
\cite{BT} and relies on the Bianchi identity satisfied by $W^I$, which
are standard super-gauge field strengths,
 \begin{equation}
  D^\al (W_{\!\al}\, \e^{-2\I V\cdot f})^I (\e^{\I V\cdot f})_I{}^J -
  \text{c.c.} = 0\ .
 \end{equation}

The vector gauge transformations generalize to
 \begin{gather}
  \de_\Lambda A^a = \Lambda^a + \bar{\Lambda}^a\ ,\quad \de_\Lambda
	V^I = 0 \\[2pt]
  \de_\Lambda \Psi_{\al I} = - \frc{\I}{4}\, \bD^2 \big[ \e^{-2\I
	V\cdot f} D_\al (\e^{\I V\cdot f} n_a \de_\Lambda A^a)
	\big]_I\ ,
 \end{gather}
where $\Lambda^a$ are arbitrary chiral superfields. For $\Psi_I$, these
have the same form as a tensor gauge transformation $\de_C$, so the
second line in \eqref{ssS} is invariant. Using the identity
\eqref{eiVT}, the transformations of $Y^a$ turn into
 \begin{equation}
  \de_\Lambda Y_\al^a = -\frc{\I}{4}\, \bD^2 \big[ \e^{-2\I V\cdot T}
  P D_\al \Lambda \big]^a + m^{aI} \de_\Lambda \Psi_{\al I}\ .
 \end{equation}
By multiplying \eqref{eiVT} from the right with $P$, one shows that
$\e^{-2\I V\cdot T}P=P$. The chirality of $\Lambda^a$ then implies that
$\bD^2D_\al\Lambda^a=0$, so we find again $\de_\Lambda Y^a=m^{aI}
\de_\Lambda\Psi_I$, which proves invariance of the action.

In the free case, $\f{IJ}{K}=0$, it is possible to add a term
 \begin{equation}
  \int\! \d^4x\, \d^2\theta\ \hat{e}_a{}^I\, \Psi_I (Y^a - \half m^{aJ}
  \Psi_J) + \text{c.c.}
 \end{equation}
to the action \eqref{ssS}, as was observed in \cite{GL}. Gauge
invariance constrains the charges $\hat{e}_a{}^I$ to satisfy \eqref{me}.
In the presence of interactions, where $\f{IJ}{K}\neq 0$, this term is
not invariant anymore under the above gauge transformations.
Unfortunately, we have not been able to find a suitable modification of
the action and transformations to restore invariance.

\begin{acknowledgments}
I would like to thank Friedemann Brandt, Bernard de~Wit and Stefan
Vandoren for stimulating discussions about dualities. Thanks also to
Mario Trigiante for making me think about massive Freedman-Townsend
models. Supported by the DFG within the priority program SPP~1096 on
string theory.
\end{acknowledgments}

\raggedright

\end{document}